\documentclass[pdftex,twocolumn,epjc3]{svjour3}          

\RequirePackage[T1]{fontenc}

\smartqed  

\RequirePackage{graphicx}
\RequirePackage{mathptmx}      
\RequirePackage{flushend}
\RequirePackage[numbers,sort&compress]{natbib}
\RequirePackage[colorlinks,citecolor=blue,urlcolor=blue,linkcolor=blue]{hyperref}

\journalname{Eur. Phys. J. C}

\usepackage{multirow}
\usepackage{amsmath}
\usepackage{hyperref}
\usepackage{amssymb}

\usepackage{soul}

\newcommand{\yl}[1]{\textcolor{black} { #1}}

\begin{document}

\title{
\yl{Catching TeV emission from GRB 221009A and alike with LHAASO, LACT and SWGO}}

\author{Yunlei Huang\thanksref{e1,addr1} \and 
        Sujie Lin\thanksref{addr1} \and
        Soebur Razzaque\thanksref{e2,addr2,addr3,addr4}\and
        Lili Yang\thanksref{e3,addr1,addr2}\and
        Zijie Huang\thanksref{addr5,addr6}
    }  


\thankstext{e1}{e-mail: huangylei3@mail2.sysu.edu.cn}
\thankstext{e2}{e-mail: srazzaque@uj.ac.za}
\thankstext{e3}{e-mail: yanglli5@mail.sysu.edu.cn}

\institute{School of Physics and Astronomy, Sun Yat-Sen University, No.2 Daxue Rd, 519082, Zhuhai China\label{addr1}
\and 
Centre for Astro-Particle Physics (CAPP) and Department of Physics, University of Johannesburg, PO Box 524, Auckland Park 2006, South Africa\label{addr2}
\and
Department of Physics, The George Washington University, Washington, DC 20052, USA\label{addr3}
\and
National Institute for Theoretical and Computational Sciences (NITheCS), Private Bag X1, Matieland, South Africa\label{addr4}
\and
University of Chinese Academy of Sciences, Beijing, 100049, China\label{addr5}
\and
Key Laboratory of Particle Astrophysics \& Experimental Physics Division \& Computing Center, Institute of High Energy Physics, Chinese Academy of Sciences, Beijing, 100049, China.\label{addr6}
}

\date{Received: date / Accepted: date}

\maketitle

\begin{abstract}
Gamma-Ray Bursts (GRBs) are the most energetic electromagnetic explosions in the universe. Recently, the Large High Altitude Air Shower Observatory (LHAASO) reported the breakthrough observation of GRB 221009A with gamma-ray energies beyond 13 TeV. This discovery, together with the previous GRB detection well above 100 GeV, confirms the production of very-high-energy (VHE, $\gtrsim 100$ GeV) radiation which might be a common component of all bright GRBs. It is reasonable to expect that bright GRBs are important targets for ground-based gamma-ray experiments. In this work, we estimate the detection rate for current and upcoming ground-based gamma-ray observatories \yl{including LHAASO, Large Array of Imaging Atmospheric Cherenkov Telescopes (LACT) and the Southern Wide-field Gamma-ray Observatory (SWGO)} under two emission models \yl{with GRB~221009A as the template}: first, that they all share the same intrinsic VHE spectral shape; second, they have the same environmental parameter and electron spectral index, governing their synchrotron self-Compton (SSC) emission. Using the \yl{long} GRB luminosity and redshift distribution function obtained from the {\em Fermi}-GBM GRB samples, and accounting for the cosmological effects and extra-galactic background light (EBL) absorption, we derive the expected VHE flux at Earth. The sensitivity analysis for LHAASO, the upcoming LACT, and SWGO to evaluate their detection potential across specific redshift and luminosity ranges has been performed. The corresponding 5$\sigma$ detection rates \yl{of 221009A-like GRBs} for the two emission models are: LHAASO, 0.04-0.05 yr$^{-1}$; LACT, 0.03-0.06 yr$^{-1}$; SWGO, 0.2-0.4 yr$^{-1}$. These rates can vary by up to $\approx 24\%$ due to different EBL models.

\PACS{
      {GRB}{}   \and
      {high-energy emission}{} \and
      {observations}{}
     } 
\end{abstract}

\section{Introduction}
\label{intro}
Gamma-ray bursts (GRBs) are the most powerful extra galactic transient events, \yl{emitting radiation across the spectrum from radio to gamma-ray bands. Their afterglows can persist from days to months after the initial prompt phase \citep{Fishman1995PASP..107.1145F}. Since their accidental discovery in 1967 by the Vela satellites \citep{Klebesadel1973ApJ...182L..85K}, they have been extensively observed and studied with space-based instruments. However, the detection of their very-high-energy (VHE, $\gtrsim 100$~GeV) radiation was long hindered by severe attenuation from the extragalactic background light (EBL) and the limited sensitivity of earlier instruments. With advances in detection technology, the} first VHE afterglow of a GRB, namely GRB~180720B was detected by H.E.S.S. on July 2018 \citep{Abdalla:2019dlr}. \yl{This was followed by the first detection of TeV emission from GRB~190114C by MAGIC in January 2019 \citep{MAGIC:2019lau}, the detection of GRB~190829A by H.E.S.S. \citep{HESS:2021dbz}, and later by MAGIC observations of GRB~201015A and GRB~201216C \citep{2020GCN.28660....1D,2020GCN.29064....1C}. These observations mark a new era in GRB studies, demonstrating the success of imaging atmospheric Cherenkov telescopes (IACTs). They have opened a new window on GRB physics, allowing VHE observations to probe extreme conditions. These observations constrain key physical parameters, such as emission mechanisms, shock properties, magnetic fields, and jet dynamics \citep[see, e.g.,][]{Gehrels2013FrPhy...8..661G, 2022Galax..10...74G}.}

Recently, the field has undergone another exciting breakthrough, where the Large High Altitude Air Shower Observatory (LHAASO) detected gamma rays from GRB 221009A up to 13 TeV \citep{LHAASO:2023kyg}. 
It was triggered by {\em Fermi}-GBM on October 9, 2022, at T$_0$ = UT 13:16:59.99 \citep{Lesage:2023vvj} and was also detected by many other space-based missions, \yl{including} Konus-Wind \citep{Frederiks:2023bxg}, AGILE \citep{2022GCN.32650....1U}, INTEGRAL \citep{2022GCN.32660....1G}, Insight-HXMT \citep{2022ATel15660....1T}, and GECAM-C \citep{Insight-HXMT:2023aof}. Detection by {\it Swift}-XRT and UVOT \citep{Williams:2023sfk} prompted extensive multi-wavelength follow-up observations \yl{that revealed a bright afterglow and measured its redshift.}

\yl{GRB~221009A ($z = 0.151$) is the brightest gamma-ray burst ever observed, with a luminosity so extreme that it saturated some detectors \citep{2022GCN.32648....1D}. Nevertheless, it has been suggested that GRB~221009A is essentially a “normal” long GRB with an exceptionally hard and energetic spectrum \citep{Frederiks:2023bxg}. Analysis of its X-ray afterglow indicates that its record-breaking brightness arises from the combination of its intrinsically high luminosity and its relatively nearby distance \citep{Williams:2023sfk}. Although such an exceptionally bright event is rare in the few-decade human observational record, it may not be uncommon over cosmological timescales. Estimates of the rate of similar energetic, nearby GRBs vary widely according to other groups, which can be from less than once per 10,000 years \citep{Burns_2023} to once per $\sim$200 years if they form a distinct local population \citep{Finke_2024arXiv240715940F}.} 


\yl{Ground-based gamma-ray detection techniques, such as extensive air shower (EAS) arrays and IACTs, have seen significant improvements recently.} Not only the energy range has been extended, but also the resolution and sensitivity have been largely enhanced. Taking the advantage of current technique and globally distributed observatories, more bright VHE sources will be revealed.
In this work, we estimate the detection rates of GRBs for two co-located observatories, LHAASO and LACT, which have overlapping energy ranges, and for the upcoming Southern Wide-field Gamma-ray Observatory (SWGO) based on its preliminary design. Our estimates are derived from the VHE gamma-ray properties of GRB~221009A and the volumetric rate of GRBs in the universe.

The paper is organized as follows. In Sec.~\ref{sec:observatories}, we compare the capabilities of current and next-generation ground-based gamma-ray observatories. The model used to predict the VHE gamma-ray spectrum of \yl{221009A-like} GRBs, as well as their luminosity and redshift distributions, is introduced in Sec.~\ref{sec:models}. Section \ref{sec:propagation} discusses the propagation effects, including cosmological redshift and absorption by the EBL. We present the sensitivity analysis of LHAASO, LACT and SWGO in Sec.~\ref{sec:obs} , followed by the corresponding detection rates in Sec.~\ref{sec:result}. Finally, our conclusions are summarized in Sec.~\ref{sec:conclusion}.

\section{Ground-based VHE gamma-ray observatories}\label{sec:observatories}
When gamma rays with energies larger than tens of GeV enter the Earth's atmosphere, they will hit the atoms there and 
\yl{dominantly produce electron-positron pairs which will further produce gamma rays through bremsstrahlung. This process continues until the energy of the electron-positron pairs falls below a critical energy of $\approx 80$ MeV. Millions of particles produced in this way forms an extensive atmospheric shower.}
To indirectly measure the primary photons on the ground, two main detection technique has been applied. One is the EAS techniques, through collecting shower particles at mountain altitudes to reconstruct the primary gamma rays \citep{cao_ultra-high-energy_2023}. Another is the Imaging Air Cherenkov (IAC) technique, by means of catching the Cherenkov light induced by the superluminous shower particles. Based on their different detection methods, they have some specific features and advantage. In recent years, a growing number of constructed and planned ground-based VHE gamma-ray observatories at diverse sites will help reveal the physical mechanisms of different TeV astrophysical sources including GRBs. 

\renewcommand{\arraystretch}{1.3}
\begin{table*}[]
\centering
\footnotesize

\begin{tabular}{|llllllll|}
\hline
\multicolumn{1}{|l|}{Observatory} & \multicolumn{1}{l|}{Location} & \multicolumn{1}{l|}{\begin{tabular}[c]{@{}l@{}}Latitude and \\ Longitude\end{tabular}} & \multicolumn{1}{l|}{Altitude} & \multicolumn{1}{l|}{\begin{tabular}[c]{@{}l@{}}Angular \\ Resolution\end{tabular}} & \multicolumn{1}{l|}{Energy Range} & \multicolumn{1}{l|}{\begin{tabular}[c]{@{}l@{}}Effective Area \\ (10TeV)\end{tabular}} & FOV \\ \hline
\multicolumn{8}{|c|}{EAS arrays} \\ \hline
\multicolumn{1}{|l|}{LHAASO} & \multicolumn{1}{l|}{Sichuan, China} & \multicolumn{1}{l|}{29.4$^{\circ}$N, 100.1$^{\circ}$E} & \multicolumn{1}{l|}{4410 m}  & \multicolumn{1}{l|}{\textgreater{}0.2$^{\circ}$} & \multicolumn{1}{l|}{100 GeV-100 PeV} & \multicolumn{1}{l|}{$5\times10^5$ m$^2$} & 1.8 sr \\ \hline
\multicolumn{1}{|l|}{SWGO} & \multicolumn{1}{l|}{Atacama, Chile} & \multicolumn{1}{l|}{23$^{\circ}$ S, 67$^{\circ}$ W} & \multicolumn{1}{l|}{4770 m}  & \multicolumn{1}{l|}{\textgreater{}0.05$^{\circ}$} & \multicolumn{1}{l|}{100 GeV-1 PeV} & \multicolumn{1}{l|}{$\sim10^6$ m$^2$} & $\sim$2 sr \\ \hline
\multicolumn{1}{|l|}{HAWC} & \multicolumn{1}{l|}{\begin{tabular}[c]{@{}l@{}}Sierra Negra, \\ Mexico\end{tabular}} & \multicolumn{1}{l|}{19.0$^{\circ}$N, 97.3$^{\circ}$W} &\multicolumn{1}{l|}{4100 m}  & \multicolumn{1}{l|}{0.1$^{\circ}$-2.0$^{\circ}$} & \multicolumn{1}{l|}{100 GeV-100 TeV} & \multicolumn{1}{l|}{$1\times10^5$ m$^2$} & 1.8 sr \\ \hline
\multicolumn{8}{|c|}{IACTs} \\ \hline
\multicolumn{1}{|l|}{LACT} & \multicolumn{1}{l|}{Sichuan, China} & \multicolumn{1}{l|}{29.4$^{\circ}$ N, 100.1$^{\circ}$ E} & \multicolumn{1}{l|}{4410 m}  & \multicolumn{1}{l|}{0.06$^{\circ}$-0.24$^{\circ}$} & \multicolumn{1}{l|}{1-300 TeV} & \multicolumn{1}{l|}{$9\times10^5$ m$^2$} & 0.02 sr \\ \hline

\multicolumn{1}{|l|}{CTAO-North} & \multicolumn{1}{l|}{La Palma, Spain} & \multicolumn{1}{l|}{28.7$^{\circ}$ N 17.9$^{\circ}$ W} & \multicolumn{1}{l|}{2200 m}  & \multicolumn{1}{l|}{\textgreater 1'} & \multicolumn{1}{l|}{20 GeV-300 TeV} & \multicolumn{1}{l|}{$8\times10^5$ m$^2$} & 0.002 sr \\ \hline
\multicolumn{1}{|l|}{CTAO-South} & \multicolumn{1}{l|}{\begin{tabular}[c]{@{}l@{}}Cerro Paranal, \\ Chile\end{tabular}} & \multicolumn{1}{l|}{24.7$^{\circ}$S, 70.3$^{\circ}$W} & \multicolumn{1}{l|}{2160 m}  & \multicolumn{1}{l|}{\textgreater 1'} & \multicolumn{1}{l|}{20 GeV-300 TeV} & \multicolumn{1}{l|}{$2\times10^6$ m$^2$} & 0.002 sr \\ \hline
\multicolumn{1}{|l|}{HADAR} & \multicolumn{1}{l|}{Tibet, China} & \multicolumn{1}{l|}{30.1$^{\circ}$N, 90.5$^{\circ}$E} & \multicolumn{1}{l|}{4300 m} & \multicolumn{1}{l|}{0.25$^{\circ}$-1.2$^{\circ}$} & \multicolumn{1}{l|}{10 GeV-10 TeV} & \multicolumn{1}{l|}{$5\times10^5$ m$^2$} & 0.84 sr \\ \hline
\end{tabular}

\caption{Basic information and performance of VHE gamma-ray observatories based on EAS or IAC techniques. See main text for references.}
\label{Table:Observatories}
\end{table*}

\subsection{EAS arrays}

Currently operating EAS detectors in the Northern Hemisphere include {High-Altitude Water Cherenkov (HAWC) Observatory} and LHAASO. To provide complementary coverage of the sky, the SWGO is planned for the Southern Hemisphere. Both HAWC and the proposed SWGO employ the water Cherenkov detector (WCD) technique, which detects Cherenkov light produced by secondary particles in water-filled tanks. LHAASO is a larger, hybrid array comprising three sub-arrays: the Water Cherenkov Detector Array (WCDA), the Square-Kilometer Array (KM2A), and the Wide Field-of-view Cherenkov Telescope Array (WFCTA). A key feature of LHAASO is the muon and electron detector in KM2A, which accurately separate the lateral distribution of electromagnetic particles and muons in the shower. This capability provides powerful background rejection, particularly at the highest energies.

Table~\ref{Table:Observatories} lists three EAS arrays that offer a nearly 100\% duty cycle and a wide field of view (FoV) exceeding 1~sr. Their angular resolution, however, is typically worse than 0.05°, and their energy threshold generally lies above 100 GeV \citep{cao_large_2022, aharonian_performance_2021, SWGO_site, doi:10.1142/9789811269776_0246, HAWC_site,collaboration_science_2025-1}. Given these characteristics, these instruments are well-suited for conducting unbiased sky monitoring and for analyzing extended sources.

\subsection{IACT arrays}

IACT arrays are usually resembled with wide optical reflectors and cameras sensitive to blue light. They detect the faint Cherenkov light emitted by secondary in an EAS, but can only operate under clear, moonless nights. This requirement limits their duty cycle to approximately 10-20\%. However, the stereoscopic reconstruction of the shower geometry observed by multiple telescopes enables a precise measurement, achieving an excellent angular resolution ($\sim1^\prime$). The large reflecting surface can improve the sensitivity at lower energies, down to tens of GeV.
Different IACT arrays vary in size and means of light acquisition, causing changes in the angular resolution ($\sim1^\prime$--$1.2^\circ$), energy threshold ($\sim 10$~GeV--1 TeV), and FOV ($\sim 0.002$--0.84~sr). Because of these features, IACT arrays are better suited for detailed studies of known sources or following-up on alerts from other wide FoV detectors.

Table.~\ref{Table:Observatories} lists four planned IACT arrays including the Cherenkov Telescope Array Observatory (CTAO)-North, CTAO-South \citep{CTAO_consortium,CTAO_site}, Large Array of Imaging Atmospheric Cherenkov Telescopes (LACT) \citep{zhang_prospects_2024-2} and High Altitude Detection of Astronomical Radiation (HADAR, \citep{qian2022prospectivestudyobservationsgammaray}), and shows their performance. CTAO consist of two arrays of telescopes, a southern-hemisphere array at ESO's Paranal Observatory and a northern array on the island of La Palma, Spain. They have three sizes of telescopes aiming at various energies of gamma rays. The LACT and HADAR experiments are both proposed gamma-ray arrays in China at LHAASO site and in Tibet, respectively. LACT is designed to consist of 32 telescopes, utilizing LHAASO muon detector array to improve the gamma-proton discrimination at higher energies. To overcome the limitation of FOV, HADAR employs wide-angle Cherenkov lens with energy range from 10 GeV to 10 TeV, and relatively poor angular resolution of 0.25$^{\circ}$ to 1.2$^{\circ}$.

The complementary characteristics of EAS particle detectors and IACT arrays lead to distinct, synergistic roles in GRB observations. EAS particle detectors, with their high duty cycle and large field of view, are well-suited for providing the trigger for transient events. In contrast, IACTs, with their superior angular resolution and higher sensitivity, are ideal for performing targeted follow-up observations. In this work, we analyze GRB detection rate with LHAASO, considering both WCDA and KM2A to provide a future prediction. Additionally, we perform a similar analysis for planed LACT, which is co-located with LHAASO, and SWGO observatories to estimate their future GRB detection potential.

\section{GRB models}\label{sec:models}

LHAASO’s observation of GRB 221009A provided the first measured spectrum of a GRB in the early afterglow phase, covering approximately $0.1-20$ TeV. The light curve observed by KM2A shows a sharp rise at $T_0+230$~s followed by a significant drop by $T_0+300$~s. Similarly, the WCDA light curve rises sharply at $T_0+230$~s, peaks around $T_0+245$~s, and then decays gradually until about $T_0+900$~s \citep{the_lhaaso_collaboration_very_2023}.

The GeV–TeV emission observed from GRBs are mainly produced by the synchrotron self-Compton (SSC) process, where electrons up-scatter their own synchrotron photons to higher energies \citep{MAGIC:2019lau, 2022Galax..10...74G, 2022Galax..10...66M, LHAASO:2023kyg, sato2023JHEAp51S, Barnard_2025}. 
To predict the intrinsic VHE spectrum of long GRBs \yl{like GRB~221009A} across different redshifts, luminosities, and \yl{the duration parameter $T_{90}$, defined as the time interval containing 5\% to 95\% of the accumulated photon counts from the source}, we adopt two emission models: a single power-law model similar to the intrinsic VHE emission spectrum of GRB 221009A derived from LHAASO data \citep{the_lhaaso_collaboration_very_2023} and a physically motivated SSC model based on Refs.~\citep{joshi_modelling_2021, Barnard:2023qkb}. Furthermore, we assume that the luminosity and redshift of long GRBs follow a specific distribution derived from the fits to the {\it Fermi}-GBM \yl{long} GRB samples \citep{banerjee_differential_2021}. The cosmological parameters throughout this work are taken from Refs.~\citep{2020A&A...641A...6P,2020Planck}.



\subsection{Power-law model}

For simplicity, we derive the intrinsic spectrum of the VHE emission by de-absorbing the observation from the brightest period, T0 + (230–300) s, for EBL attenuation using the model given by \citep{saldana-lopez_observational_2021-1}. This de-absorbed spectrum serves as a template spectral shape for all GRBs in our simulation.
Fitted with a simple power law, the spectrum has a normalization of $144\pm 4 \times 10^{-8}$~TeV$^{-1}$~cm$^{-2}$~s$^{-1}$ and a spectral index of $-2.35\pm 0.03$ \citep{the_lhaaso_collaboration_very_2023}.

Based on this intrinsic spectrum, we derive the VHE luminosity $L_0(\nu)$ of GRB~221009A as:
\begin{equation}
    L_0(\nu) = 4\pi D_L^2(z_0) F \left(\frac{\nu}{1+z_0} \right),
\end{equation}
where $z_0$ is the redshift of GRB 221009A, $F(\nu/(1+z_0))$ is the EBL-corrected observed flux during $T_0+(230-300)$~s interval transformed to the source rest frame, and $D_L(z_0)$ is the luminosity distance corresponding to $z_0$ \citep{refId0}.
The intrinsic VHE luminosity $L_i(\nu)$ of a generic GRB labeled $i$ is then obtain by scaling $L_0(\nu)$ proportionally to its isotropic-equivalent  $E_{i,\rm iso}$ in the $(1-10^4)$ keV band to that of GRB 221009A, $E_{0,\rm iso} \approx 10^{55}$ erg \citep{Lesage:2023vvj}:
\begin{equation}
    L_i(\nu) = \frac{E_{i,\rm iso}}{E_{0,\rm iso}} L_0(\nu) \,.
\end{equation}
Here, $E_{i,\rm iso}$ is derived from the peak luminosity $L_{i,\rm pk}$ in the same energy band as $E_{i,\rm iso} = T_{90} \times (1+z) \times L_{i,\rm pk}$. For the prompt-emission duration $T_{90}$ of long GRBs, we adopt a log-normal distribution based on {\em Fermi}-GBM observations \citep{von_kienlin_fourth_2020}.

\subsection{SSC model}

To predict the VHE emission of \yl{221009A-like GRBs}, we adopt the GRB afterglow synchrotron+SSC model developed in~\citep{joshi_modelling_2021, Barnard:2023qkb}, which has successfully described the spectra of several VHE GRBs. 
The model is characterized by six parameters in the slow-cooling spectrum for the afterglow in the ISM scenario: the time $t$ after the trigger, the isotropic-equivalent initial kinetic energy $E_k$, the interstellar medium (ISM) number density $n$, the electron energy distribution index $p$, and the fractions of shock energy carried by electrons and magnetic fields, $\epsilon_e$ and $\epsilon_B$, respectively.

{To better constrain the parameters of the SSC model, we incorporated the MeV–GeV observations of GRB~221009A from the AGILE satellite \citep{tavani_agile_2023} in addition to the LHAASO data. AGILE detected the burst from the prompt phase into the early afterglow, covering the intense emission between 200~s and 300~s post‑trigger and revealing persistent, non-thermal GeV radiation lasting up to $\sim 10^4$~s. These data provide essential spectral and flux information that bridges the keV prompt emission and the later TeV afterglow.}

We use the VHE spectra of GRB 221009A from two time intervals, $T_0 + [230, 300]$~s and $T_0 + [300, 900]$~s \citep{the_lhaaso_collaboration_very_2023} {together with the data collected by AGILE within a similar time period of $T_0 + [248, 326]$~s and $T_0 + [326, 900]$~s \citep{Foffano_2024}}, to constrain these parameters. The onset of the afterglow is taken as $T^* = T_0 + 226$~s \citep{LHAASO:2023kyg}.
The representative time for each spectral window is defined around its midpoint, corresponding to $t = 49$~s and $t = 374$~s after $T^*$ for the first and second intervals, respectively.

For the fitting, we fixed $E_k = \eta E_{\rm iso} =1 \times10^{55}$~erg, where the factor $\eta =1$, as physically motivated assumptions, while treating $p$, $\epsilon_e$ and $\epsilon_B$ as free parameters. These are fit simultaneously to the two spectra using a Markov Chain Monte Carlo (MCMC) method \citep{Foreman_Mackey_2013}. {Assuming that these three parameters remain constant across the two time intervals, we perform a simultaneous fit to both spectra by combining their likelihoods, yielding the following result:}: $p = 2.05$,  $\epsilon_e = 0.002$, $\epsilon_B = 0.0006$. {The details of the fitting can be found in \ref{app2}.} 

To generate the spectra of GRBs with different peak luminosity $L_{i,\rm pk}$, redshift $z$ and prompt duration $T_{90}$, the initial kinetic energy is estimated as: 
\begin{equation}
    E_{i,\rm k} = \eta \times T_{90} \times L_{i,\rm pk}  (1+z)\,,
    \label{eq:Eik}
\end{equation}
where $E_{i,\rm iso} \approx T_{90}\,L_{i,\rm pk} (1+z)$ and we assume $\eta = 1$ as a conservative value guided by the fit to GRB 221009A. The density $n$, spectra index $p$, and energy fractions $\epsilon_e$ and $\epsilon_B$ are held fixed at the values obtained from the fit to GRB 221009A.  Under this assumption, the time‑dependent SSC model can be used to construct the intrinsic VHE spectral evolution for each GRB.

\subsection{GRB distribution}
\label{subsec:grbdist}
The luminosity and redshift distribution of long GRBs, $N(L,z)$, is obtained from the fit to the {\em Fermi}-GBM GRB sample presented in~\citep{banerjee_differential_2021}. This distribution is constrained by the differential peak‑flux counts via the integral:
\begin{equation}
    N(>F)  =  \int_{L_{\rm min}}^{L_{\rm max}} \int_0^{z(L,F)} N(L,z) \,\frac{dV_{\rm com}(z)}{dz} \, \frac{1}{1+z} \,dz \,dL \,.
    \label{eq.integration_num}
\end{equation}
where $L_{\rm min}$ and $ L_{\rm max}$ are the lower and upper luminosity limits, and $z(L,F)$ is the maximum redshift at which a burst of luminosity $L$ would produce an observed peak flux $F$. The term $dV_{\rm com}(z) / dz$ denotes the comoving volume element per unit redshift, and the factor $(1+z)^{-1}$ accounts for cosmological time dilation. Among the physical models considered in~\citep{banerjee_differential_2021}, we employ the “scale‑free distribution of blind‑spot sizes”, which can be expressed as:
\begin{align}
\begin{split}
    N(L,z) & =  c_0 n(L)\, \rho_0 \rho(z) \\
    & = c_0\, \rho_0 \left[\left(\frac{L}{L_{0}}\right)^{-\alpha} + \epsilon\left(\frac{L}{L_{0}} \right)^{-\beta} \right] \\
    &~~~~\times \frac{(1+z)^{2.7}}{1+[(1+z)/2.9]^{5.6}}
    \label{eq.distribution}
\end{split}
\end{align}
where $n(L)$ is the peak-luminosity function fitted from the \yl{long} GRBs detected by {\em Fermi}-GBM (corresponding to $L_{i,\rm pk}$), $c_0$ is its normalization, $\rho(z)$ is the star formation rate \citep{2014ARA&A..52..415M} and $\rho_0 \approx 5.5$~Gpc$^{-3}$yr$^{-1}$ is the local GRB rate.
The remaining parameters, obtained from~\citep{banerjee_differential_2021}, are $\alpha = 1.33$, $\beta = 1.42$, $\epsilon =10$, and $L_{0} = 3 \times 10^{53}$~erg~s$^{-1}$ \citep{banerjee_differential_2021}.

The parameter space is sampled with 80 linear points in redshift $z \in [0, 2]$, 80 logarithmic points in peak luminosity $L_{i,\rm pk}\in[10^{46},10^{54}]$~erg~s$^{-1}$, and 30 points in $T_{90}$ within $\pm4\sigma$ of its log-normal distribution. For each of the resulting synthetic GRBs $80 \times 80 \times 30$, we generate the intrinsic spectrum using the two emission models described above. The chosen luminosity range covers both high- and low-luminosity GRB populations \citep{irwin_jet_2016-1, 2007A&A...465....1D}, and the redshift is limited to $z\le2$ as TeV emission is strongly attenuated by the EBL at higher redshifts.

\section{Propagation of VHE gamma-rays}\label{sec:propagation}

The observed VHE gamma-ray spectrum from a GRB is modified by two primary propagation effects: cosmological redshift and attenuation due to pair production ($\gamma\gamma\to e^\pm$) with EBL photons. We account for both processes when transforming the intrinsic spectrum to the observed one at Earth.

\subsection{Redshift effect}
GRBs are cosmological events, often located at redshifts of $z \sim 1$–$2$. Consequently, the redshift effect significantly influences their observed spectra. As photons propagate through the expanding universe, their energies are redshifted. For an intrinsic spectral energy range $E_1 \leq E \leq E_2$, the observed differential flux at redshift $z$ is given by:
\begin{equation}
    F(z) = \frac{\tilde{L}(z)}{4\pi D_L(z)^2} ,
\end{equation}
where $\tilde{L}$ is the spectral luminosity (in erg s$^{-1}$) evaluated in the rest-frame energy interval $E_1(1+z) \leq E \leq E_2(1+z)$, and $D_L(z)$ is the luminosity distance \citep{refId0}. This results in an overall shift of the spectrum to lower energies by a factor of $(1+z)$. Because the intrinsic spectrum of GRB 221009A in the interval $T_0 + [230, 300]$~s shows no evidence of a cutoff up to the highest observed energies, we adopt 20~TeV as the maximum intrinsic photon energy for all redshifts considered in our calculations.

\subsection{EBL absorption}

The EBL consists of diffuse radiation spanning the near-infrared to ultraviolet wavelengths, originating from the emitting light of star formation processes throughout the history of the universe \citep{Razzaque2009ApJ...697..483R}. The spectral energy distribution of the EBL evolves with redshift. VHE gamma-ray propagating over cosmological distances are primarily attenuated via pair production ($\gamma\gamma\to e^\pm$) on the infrared–optical photons of the EBL.

The resulting attenuation is quantified by an energy- and redshift-dependent optical depth, $\tau_\gamma(E, z)$, where $E$ is the observed photon energy and $z$ is the source redshift. The observed flux after EBL absorption is related to the intrinsic spectrum (already corrected for redshift) by:

\begin{equation}
    F_{\rm obs}(E) = F_{\rm in}(E)\, e^{-\tau_\gamma(E,z)}
\end{equation}
where $F_{\rm in}(E)$ is the intrinsic flux. The optical depth $\tau_\gamma$ increases steeply with energy in the TeV range and is model-dependent. Several EBL models, constructed using different astrophysical and empirical approaches, provide estimates for $\tau_\gamma(E,z)$ \citep[e.g.,][]{franceschini_extragalactic_2008, Razzaque2009ApJ...697..483R, Finke:2009xi, dominguez_extragalactic_2011, gilmore_semi-analytic_2012}. In this work, we adopt the model in~\citep{saldana-lopez_observational_2021-1}, which derives the EBL evolution out to $z \sim 6$ based on observed galaxy data.

To illustrate the level of attenuation and the associated model uncertainties, we consider GRB 221009A at $z = 0.151$. For the model of Saldana-Lopez et al.~(2021)~\citep{saldana-lopez_observational_2021-1}, the gamma-ray survival fraction (i.e., $e^{-\tau_\gamma}$) is approximately 20\% at 1~TeV and 0.1\% at 10~TeV. Other EBL models yield different values; for instance, at 1~TeV the survival fraction can be as high as 18\% \citep{Finke:2009xi}, and at 10~TeV it may reach 0.37\% \citep{gilmore_semi-analytic_2012}. The impact of these EBL model differences on the predicted GRB detection rates will be discussed in Section~\ref{sec:Rate}.

\section{Analysis to detect VHE GRBs}\label{sec:obs}

We estimate the detectability and detection rates of \yl{221009A-like} GRBs at VHE by computing the expected number of signal and background events for a given flux model. The statistical significance of a detection is then evaluated using a test statistic (TS) derived.

\subsection{Expected Event Counts}

For a GRB with peak luminosity $L$, redshift $z$, and duration $T_{90}$, the observed differential VHE flux $F(E)$ is obtained by applying the intrinsic spectral models (Sec.~\ref{sec:models}) and the propagation effects (Sec.~\ref{sec:propagation}). The expected number of events in an energy bin $[E_{\rm min, i} \cdot E_{\rm max, i}]$ from both the source ($N_{\rm src}$) and the background ($N_{\rm bkg}$) is calculated as:
\begin{equation}
    N(E_{\rm i}) = \int_{E_{\rm min}}^{E_{\rm max}} \int_{t_i}^{t_f} \epsilon(E)\, A_{\rm eff}(E)\, F(E,t)\, dtdE,
    \label{count}
\end{equation}
where $\epsilon(E)$ is the survival fraction {of gamma rays} after $p/\gamma$ discrimination, $A_{\rm eff}(E)$ is the instrument’s effective area, assumed here to be identical for gamma rays and protons for the purpose of background estimation, and $F(E,t)$ is the time-dependent differential flux.

For the simple power-law emission model, we use the spectrum averaged over the brightest phase, $T_0+[230-300]$~s, and assume it remains constant over this interval, so the time integral reduces to a multiplicative factor of $\Delta T = 70$~s. For the physically motivated SSC model, the flux $F(E,t)$ is explicitly time-dependent. We integrate from $t_i = T_{90} / 2$ ({proxy for the peak flux time, when the afterglow starts}) to $t_f = 3000$~s to encompass the bulk of the detectable afterglow emission.

The effective areas $A_{\rm eff}(E)$, energy bins, and values of $\epsilon(E)$ used for LHAASO, LACT, and SWGO are provided in \ref{app1}. For LACT, since only the effective area of the current 8‑IACT array is available, we scale it by a factor of four to approximate the performance of the planned configuration with 32 IACTs \citep{zhang_layout_2025}.

\subsection{Sensitivity analysis}\label{sec:sens}

We assess the detectability of each GRB using the ON/OFF method and the likelihood ratio test. In this framework, the background count in a given region is taken as $N_{\rm off} = N_{\rm bkg}$, and the total count in the source region is $N_{\rm on} = N_{\rm bkg} + N_{\rm src}$. The {TS} is computed via the Li-Ma formula \citep{li_analysis_1983}:
\begin{equation}
\begin{split}
    TS = & -2 
    \left[ \sum_{i=1}^n N_{\rm  on}^i \ln\left(\frac{\alpha}{1+\alpha} \frac{N_{\rm on}^i + N_{\rm off}^i}{N_{\rm on}^i}\right) \right. \\
    & \left. + ~N_{\rm off}^i\, \ln\left(\frac{1}{1+\alpha} \frac{N_{\rm on}^i + N_{\rm off}^i}{N_{\rm off}^i}\right) \right]
\end{split}
\end{equation}
where the on‑source and off‑source exposure times are equal ($\alpha=1$), and the sum runs over the $n$ energy bins. For each simulated GRB which defined by its luminosity and redshift, we obtain a corresponding TS value. Roughly a TS $= 25$ corresponds to a detection significance {with} $5\sigma$ and TS $= 9$ with $3\sigma$.

\begin{figure*}[htbp]
  \centering
  \resizebox{1.0\textwidth}{!}{%
  \includegraphics{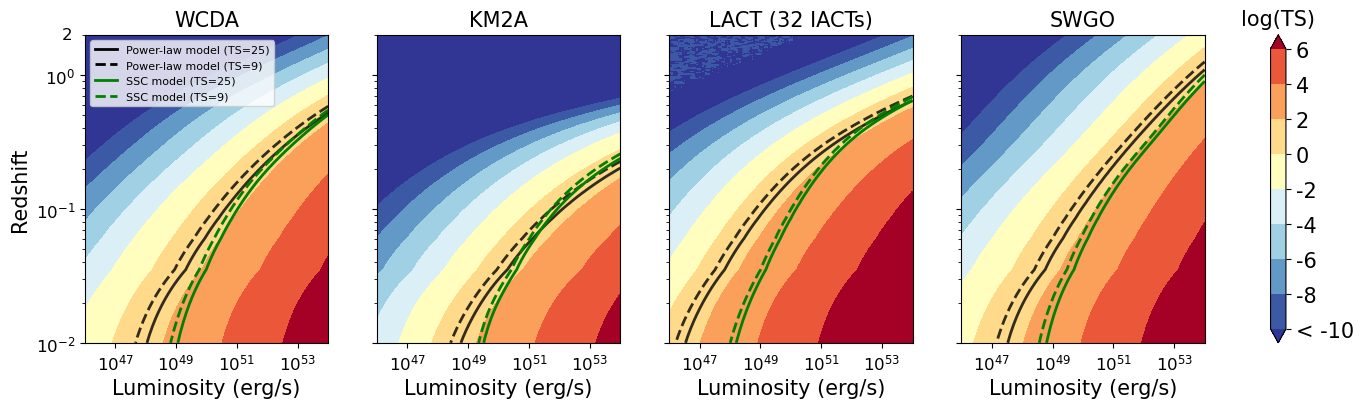}}

  \caption[]
  {\label{TS_map}{The contours of $\log(TS)$ for \yl{221009A-like} GRBs across a range of luminosities and redshifts, calculated for LHAASO‑WCDA, LHAASO‑KM2A, LACT, and SWGO. All panels assume the most probable prompt duration $T_{90} = 28$~s. and a 68\% containment of the point‑spread function. Solid and dashed black lines mark the TS = 25 and TS = 9 boundaries for the power‑law model, corresponding to $5\sigma$ and $3\sigma$ detection regions, respectively. Similarly, solid and dashed green lines indicate the corresponding boundaries for the SSC model. See the main text for further details.
  }}
\end{figure*}

Figure~\ref{TS_map} presents the TS contours for LHAASO (WCDA and KM2A), LACT, and SWGO, showing how the detectability thresholds of $TS=25$ and $TS=9$ varies with GRB luminosity and redshift for both the power‑law and SSC emission models.

\section{Results and Discussion}\label{sec:result}

\begin{figure*}
  \centering
    \resizebox{0.75\textwidth}{!}{%
      \includegraphics{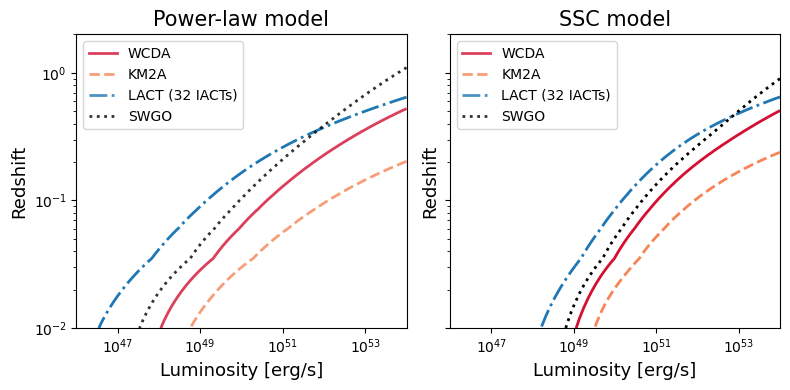}}
  \caption[]
    {\label{TS_line}
    The detectability of VHE emission from \yl{221009A-like} GRBs at the $5\sigma$ level is shown in the redshift–luminosity plane for LACT, WCDA, KM2A, and SWGO, assuming the most probable prompt duration $T_{90} = 28$~s. Here, the luminosity represents the peak luminosity in the 1-$10^4$ keV band for each simulated GRB. The red solid, orange dashed, blue dash‑dotted, and black dotted curves correspond to LHAASO‑WCDA, LHAASO‑KM2A, LACT (with 32 IACTs), and SWGO, respectively. For each experiment, the detectability is evaluated based on the expected number of events within the 68\% containment radius of the point-spread function.
    }
\end{figure*}

\begin{table*}[ht]
\renewcommand{\arraystretch}{1.5}
\centering
\footnotesize

\begin{tabular}{llllll}
\hline
Model name & Detector & WCDA & KM2A & LACT & SWGO \\ \hline
\multicolumn{1}{l|}{\multirow{2}{*}{\begin{tabular}[c]{@{}l@{}}Power-law model\end{tabular}}} & $5\sigma$ Detection Rate (yr$^{-1}$) & $0.050_{-0.003}^{+0.003}$ & $0.0037_{-0.0002}^{+0.0002}$ & $0.059_{-0.003}^{+0.003}$ & $0.420_{-0.030}^{+0.028}$ \\ \cline{2-6} 
\multicolumn{1}{l|}{} & $3\sigma$ Detection Rate (yr$^{-1}$)& $0.078_{-0.005}^{+0.005}$ & $0.0060_{-0.0003}^{+0.0003}$ & $0.085_{-0.004}^{+0.004}$ & $0.658_{-0.044}^{+0.043}$ \\ \hline

\multicolumn{1}{l|}{\multirow{2}{*}{SSC model}} & $5\sigma$ Detection Rate (yr$^{-1}$) & $0.035_{-0.001}^{+0.001}$ & $0.0047_{-0.0001}^{+0.0001}$ & $0.034_{-0.001}^{+0.001}$ & $0.175_{-0.004}^{+0.004}$ \\ \cline{2-6} 
\multicolumn{1}{l|}{} & $3\sigma$ Detection Rate (yr$^{-1}$) & $0.047_{-0.001}^{+0.001}$ & $0.0064_{-0.0001}^{+0.0001}$ & $0.043_{-0.001}^{+0.001}$ & $0.247_{-0.006}^{+0.005}$\\ \hline
\end{tabular}

\caption{Predicted annual \yl{221009A-like} GRB detection rates for LHAASO (WCDA and KM2A), LACT, and SWGO, calculated using the EBL model of Saldana-Lopez et al.~(2021) \citep{saldana-lopez_observational_2021-1}. The quoted uncertainties incorporate the $1\sigma$ flux uncertainty for the power-law model and a 7\% systematic uncertainty common to both the power-law and SSC models. 
}

\label{Table:detection rate}
\end{table*}

\normalsize

\subsection{Detectability of VHE \yl{221009A-like} GRBs}
\yl{Figure \ref{TS_map} shows the log(TS) contours from the sensitivity analysis for 221009A-like GRBs across a range of luminosity and redshift. GRBs with higher luminosities and closer distances are clearly more detectable for each detectors. The 3$\sigma$ detection region is not substantially larger than the 5$\sigma$ region. Compared to the SSC model, the power-law model predicts significantly higher TS values at low luminosity and low redshift, especially for LACT, while the two models converge at higher luminosities and redshifts. This occurs because the SSC model predicts a lower afterglow flux for less energetic GRBs, making the power-law model more optimistic in that regime.}

\yl{Interestingly, for high-luminosity GRBs, the SSC model predicts TS values that gradually approach and eventually exceed those of the power-law model as the energy threshold of the instrument increase—specifically, from SWGO (which has the lowest energy threshold and band) to WCDA, LACT, and KM2A (with progressively higher energy thresholds). This behavior may be related to our choice of integration time: the SSC model includes afterglow emission lasting up to 3000~s, which can contribute significantly to the VHE signal for very luminous bursts.}

Figure \ref{TS_line} shows the $TS=25$ detectability contours for WCDA, KM2A, LACT, and SWGO in the redshift–luminosity ($1-10^4$~keV) plane, for both the power‑law and SSC emission models. \st{In general, the detectable region predicted by the power‑law model is larger than that of the SSC model for all instruments except KM2A.} \yl{As shown in the figure, at} a redshift $z\approx 0.3$, the minimum detectable peak luminosity is roughly $L \gtrsim (3-5)\times10^{52}$~erg~s$^{-1}$ for WCDA, $L \gtrsim (2-4)\times10^{51}$~erg~s$^{-1}$ for LACT, and $L \gtrsim (0.3-1)\times10^{52}$~erg~s$^{-1}$ for SWGO. GRBs become difficult to detect with KM2A at redshifts beyond $z \gtrsim 0.2$, and for all four experiments beyond redshifts $z \gtrsim 1$. Comparing SWGO and LACT, SWGO performs better for high‑luminosity GRBs \yl{due to the lower energy threshold}, whereas LACT retains an advantage for low‑luminosity sources owing to its superior sensitivity in that regime. \yl{According to the power-law model, both LACT and SWGO could detect low-luminosity GRBs (<$10^{48}$~erg~s$^{-1}$), whereas the SSC model cannot.}

\subsection{Detection rate}\label{sec:Rate}
Having examined the detectability of VHE gamma-ray emission from GRBs for LHAASO, LACT, and SWGO at different significance levels, we now compute their expected detection rates.

The $5\sigma$ and $3\sigma$ visible region for a given instrument and a fixed $T_{90}$, {(chosen to be 28~s, the peak of the $T_{90}$ distribution \citep{von_kienlin_fourth_2020}, for illustration)} is defined by the boundary $z(L)$ at which the test statistic $TS\ge 25$ and $TS\ge 9$ respectively, as shown in Fig.~\ref{TS_map} and Fig.~\ref{TS_line}, respectively. For each $T_{90}$, the corresponding partial detection rate $R_{T_{90}}$ is obtained by integrating the GRB population model $N(L,z)$ (described in Sec.~\ref{subsec:grbdist} and fitted to the \textit{Fermi}-GBM data) over the visible region. This is performed by evaluating Eq.~(\ref{eq.integration_num}) with the flux limit $z(L,F)$ replaced by the detectability boundary $z(L)$. 

The total rate $R_{\rm tot}$ for each instrument is then calculated by summing over all $R_{T_{90}}$ values, weighted by the log‑normal probability distribution of $T_{90}$ \citep{von_kienlin_fourth_2020} introduced in Sec.~\ref{sec:models}. Finally, the instrument‑specific annual detection rate is given by:
\begin{equation}
    \mathrm{Rate} = R_{\mathrm{tot}} \times f_{\mathrm{FOV}} \times f_{\mathrm{duty}},
\end{equation}
where $f_{\mathrm{FOV}}$ is the fractional FoV and $f_{\mathrm{duty}}$ is the duty cycle. For LHAASO and SWGO, the duty cycle is nearly 100\%, while for LACT we adopt 18\%. The corresponding FoV fractions are listed in Table~\ref{Table:Observatories}. {For SWGO, we assume a zenith-angle coverage of 0$^\circ$–52$^\circ$ \citep{collaboration_science_2025-1}.} Because LACT can slew to follow up external alerts, we assume it can observe any source within a zenith angle of $60^\circ$, which defines its effective sky coverage.

\begin{table*}[]
\renewcommand{\arraystretch}{1.5}
\centering
\footnotesize

\begin{tabular}{lllllll}
\hline
Model name & Significance & EBL model & WCDA & KM2A & LACT & SWGO \\ \hline
\multirow{4}{*}{Power-law model} & \multirow{2}{*}{\begin{tabular}[c]{@{}l@{}}5-sigma detection\\ rate (yr$^{-1}$)\end{tabular}}  & Finke et al.~(2010) \citep{Finke:2009xi} 
 & 0.042  & 0.0035 & 0.048 & 0.367 \\ \cline{3-7} 
 &  & Gilmore et al.~(2012) \citep{gilmore_semi-analytic_2012} 
 & 0.056 & 0.0046 & 0.065 & 0.467 \\ \cline{2-7} 
 & \multirow{2}{*}{\begin{tabular}[c]{@{}l@{}}3-sigma detection\\ rate (yr$^{-1}$)\end{tabular}}  & Finke et al.~(2010)  \citep{Finke:2009xi} 
 & 0.066 & 0.0054 & 0.069 & 0.567 \\ \cline{3-7} 
 &  & Gilmore et al.~(2012) \citep{gilmore_semi-analytic_2012} 
 & 0.088 & 0.0073 & 0.094 & 0.726 \\ \hline
\multirow{4}{*}{SSC model} &\multirow{2}{*}{\begin{tabular}[c]{@{}l@{}}5-sigma detection\\ rate (yr$^{-1}$)\end{tabular}}  & Finke et al.~(2010) \citep{Finke:2009xi}
& 0.029 & 0.0042 & 0.027 & 0.153 \\ \cline{3-7} 
 &  & Gilmore et al.~(2012) \citep{gilmore_semi-analytic_2012} 
 & 0.039 & 0.0058 & 0.038 & 0.197 \\ \cline{2-7} 
 & \multirow{2}{*}{\begin{tabular}[c]{@{}l@{}}3-sigma detection\\ rate (yr$^{-1}$)\end{tabular}} & Finke et al.~(2010) \citep{Finke:2009xi}
 & 0.039 & 0.0055 & 0.034 & 0.214 \\ \cline{3-7} 
 &  & Gilmore et al.~(2012) \citep{gilmore_semi-analytic_2012} 
 & 0.054 & 0.0079 & 0.048 & 0.277 \\ \hline
\end{tabular}

\caption{Comparison of predicted \yl{221009A-like} GRB detection rates for LHAASO (WCDA and KM2A), LACT, and SWGO under different EBL models \citep{Finke:2009xi, gilmore_semi-analytic_2012} and emission models (power-law and SSC).}
\label{Table:diff EBL}
 
\end{table*}

Table \ref{Table:detection rate} presents the predicted annual detection rates for \st{long} \yl{221009A-like} GRBs at $5\sigma$ and $3\sigma$ significance for LHAASO, LACT, and SWGO\st{, based on the GRB luminosity and redshift distribution inferred from {\em Fermi}-GBM observations} under two emission models. The uncertainties quoted for the power-law model incorporate the normalization error of the fitted intrinsic spectrum and an additional 7\% systematic uncertainty. For the SSC model, only the 7\% systematic uncertainty is included.
{For all detectors}, the detection rates predicted by the simpler power-law model are systematically {higher} than those from the SSC model. For LHAASO, the $5\sigma$ detection rate is approximately 0.04-0.05~yr$^{-1}$ for WCDA and about 0.004-0.005~yr$^{-1}$ for KM2A. SWGO shows the highest rate, around 0.2-0.4~yr$^{-1}$. LACT is predicted to detect 0.03-0.06~yr$^{-1}$ at $5\sigma$, limited by its lower duty cycle ($\sim 18\%$) and higher energy threshold, where EBL absorption between 1 and 10 TeV is strong. At the $3\sigma$ level, 
all the rates increases, the highest being 0.7~yr$^{-1}$ for SWGO.
%
These rates imply that LHAASO-WCDA may detect one GRB roughly every $\approx 20$ years, while KM2A might require $\approx 210-270$ years per detection. The choice of EBL model also influences the predicted rates by altering the observed spectral shape and flux. As shown in Table~\ref{Table:diff EBL}, using the model of Finke et al.~(2010) \citep{Finke:2009xi} yields lower rates for all instruments, whereas the model of Gilmore et al.~(2012) \citep{gilmore_semi-analytic_2012} gives higher rates; the relative deviation between {different EBL models can reach 24\%.} 

\subsection{Comparison with other detectors}
{Beyond the experiments studied in this work, CTAO represents a major upcoming facility for ground-based gamma-ray astronomy. Employing the IAC technique, CTAO is projected to achieve unprecedented sensitivity in the TeV band, with predicted GRB detection rates ranging from approximately one event every 20–30 months \citep{gilmore_iact_2013-1} to more optimistic estimates exceeding 1~yr$^{-1}$ \citep{ CTAO_consortium}.}

{For the existing IACTs, such as H.E.S.S., MAGIC, and VERITAS, which have already successfully detected several VHE GRBs, the rate of GRBs detectable by them is estimated to be $<1$~yr$^{-1}$ based on their current performance \citep{ashkar_case_2024}. This estimate does not account for further reductions due to adverse weather, hardware downtime, or latency in automated response to external alerts, all of which could significantly affect final detection statistics. \yl{Consequently, the predicted rates presented here are lower than what would be achieved in actual observational campaigns.}
}

{For HAWC observatory, an EAS array currently operational, is estimated to detect $\sim$ \yl{0.01-0.25 long} GRBs per year \yl{based on the sensitivity analysis for different cut-off energy (100-500GeV) of GRBs. \citep{TABOADA2014276}}


In summary, the VHE GRB detection rate for any observatory is strongly influenced by its duty cycle, field of view, and energy threshold. EAS arrays (like LHAASO and SWGO) offer high duty cycles and wide fields of view but operate above $\sim 100$~GeV. IACTs (like LACT and CTAO) provide better sensitivity and lower energy thresholds ($\sim 10$~GeV) but are limited by lower duty cycles. The co-located LHAASO and LACT can effectively work in tandem, with LHAASO providing triggers for LACT follow-up observations across 0.1-400~TeV. In the future, SWGO will complement LHAASO by monitoring the southern sky \citep{hinton_southern_2021}, and planned instruments like HADAR—with a large field of view (0.84~sr) and a low energy threshold ($\sim 10$~GeV) based on an innovative water-lens imaging atmospheric Cherenkov technique \citep{qian2022prospectivestudyobservationsgammaray}, which could further enhance GRB detection capabilities.

\section{Conclusions}\label{sec:conclusion}
The detection of the early TeV afterglow from GRB 221009A marked a breakthrough in VHE gamma-ray astronomy. This event, together with earlier GRB detections well above 100 GeV, suggests that VHE emission may be a common feature of bright GRBs. Identifying the next luminous VHE GRB therefore represents a key objective for current and upcoming ground-based observatories such as LHAASO, LACT and SWGO which we have focused on here.

In this work, we have adopted two complementary approaches to model the intrinsic VHE spectrum: a simple power-law model, in which all long GRBs are assumed to share the same spectral shape as GRB~221009A, with their luminosities scaled according to the ratio of their isotropic-equivalent energy release in the $1-10^4$ keV band; and a physically motivated SSC model, in which the ISM density, electron energy distribution index, and the fractions of shock energy carried by electrons and magnetic fields are fixed to the values inferred for GRB 221009A, while the initial kinetic energy is scaled. In both cases, the GRB population is simulated using the peak‑luminosity, redshift, and $T_{90}$ distributions derived from the {\it Fermi}-GBM catalogue, enabling a robust estimate of the expected VHE detection rates.
Our main results can be summarized as follows:

\begin{itemize}

\item \textbf{Detectability ranges:} For \yl{221009A-like} GRBs with peak luminosities {($1-10^4$ keV band)} in the range $10^{50} \lesssim L_{\mathrm{peak}} \lesssim 10^{54}\,\mathrm{erg~s^{-1}}$, WCDA can detect sources within a redshift interval of $z \lesssim (0.04-0.5)$ at the $5\sigma$ significance level. Over the same luminosity interval, the detectable redshift ranges are $z \lesssim (0.02-0.2)$ for KM2A, $z \lesssim (0.09-0.7)$ for LACT, and $z \lesssim (0.05-1.1)$ for SWGO, all at $5\sigma$ significance.

\item \textbf{Detection rates:} The predicted $5\sigma$ detection rates vary due to different EBL models {and VHE emission models}, yielding approximately 0.04-0.05~yr$^{-1}$ for LHAASO‑WCDA, 0.004-0.005~yr$^{-1}$ for LHAASO-KM2A, 0.03-0.06~yr$^{-1}$ for LACT, and 0.2-0.4~yr$^{-1}$ for SWGO. The comparable rates of WCDA and LACT highlight how the low duty cycle of IACTs can strongly limit their detection efficiency for transients, even when their point-source sensitivity is superior to that of EAS arrays. SWGO’s higher rate stems from its larger effective area and improved $p/\gamma$ discrimination.

\item \textbf{Instrumental drivers:} The detection rate of a TeV observatory is governed primarily by its duty cycle, field of view, and energy threshold. A lower energy threshold reduces the impact of EBL attenuation and generally enhances the detection prospect. SWGO’s higher rate illustrates that an EAS array with a lower threshold, larger effective area, and better background rejection can significantly improve the monitoring and detection of VHE GRBs.

\item \textbf{EBL model dependence:} Although EBL absorption strongly suppresses the observed TeV flux, the resulting variation in predicted detection rates across different EBL models remains modest, not exceeding $\approx $\(24\%\).
\end{itemize}

Our analysis indicates the detection of GRB~221009A by KM2A was a particularly fortunate event, occurring near the instrument’s redshift sensitivity limit. We estimate that another similarly bright, nearby GRB might be detected by KM2A only once every $\sim$ 210-270 years. Less energetic or higher‑redshift GRBs will be detected more frequently. SWGO is expected to achieve a $5\sigma$ detection rate of about 0.2-0.5~yr$^{-1}$ {depending on EBL model}s, which is {slightly} lower than the SWGO collaboration’s own estimate of $\sim$ 0.7 yr$^{-1}$ \citep{collaboration_science_2025-1} due to our more conservative assumptions regarding the intrinsic VHE spectrum.

In summary, the high duty cycle and wide field of view of LHAASO and the future SWGO make them ideal instruments for providing triggers for follow‑up observations by more sensitive, pointed instruments such as LACT and CTAO. The eventual detection of a large sample of VHE GRBs will be essential for probing the physical origin of TeV emission, constraining jet structures and circumburst environments, and understanding particle acceleration in relativistic shocks. Moreover, the synergy between ground‑based gamma-ray observatories and other multi‑messenger facilities, such as gravitational wave detectors and neutrino telescopes, promises to open new windows into the most energetic processes in the Universe.

\begin{acknowledgements}
The authors thank Min Zha for the helpful discussion. SR acknowledges partial supports from the National Research Foundation (NRF), South Africa, BRICS STI programme; and the National Institute for Theoretical and Computational Sciences (NITheCS), South Africa. LY acknowledges supports from the National Natural Science Foundation of China (NSFC), Grants No.\ 12261141691 and No.\ 12005313 and the support from Fundamental Research Funds for the Central Universities, Sun Yat-sen University, No. 24qnpy123.
\end{acknowledgements}

\bibliographystyle{elsarticle-num}
\bibliography{ref}











\appendix

\section{{SSC model fitting to GRB 221009A}}\label{app2}

The Markov Chain Monte Carlo (MCMC) fitting procedure employed in this work utilizes the \texttt{emcee} package \citep{Foreman_Mackey_2013} to constrain the parameters of the synchrotron self-Compton (SSC) model applied to the GRB afterglow spectral energy distribution (SED). Observational data of GRB 221009A from LHAASO and AGILE are combined across two time intervals: $T^*+[4,74]$~s (interval 1) and $T^*+[74,674]$~s (interval 2) \citep{the_lhaaso_collaboration_very_2023}. Even if the time periods corresponding to the AGILE spectrum do not exactly align with the settings of LHAASO ($T^*+[22,100]s$ and $T^*+[100,674]s$), we can roughly consider it as radiation emitted within a certain time period \citep{Foffano_2024}. The representative time for each spectral window is defined as around its midpoint, corresponding to $t = 49$~s and $t = 374$~s after $T^*$ for the first and second intervals, respectively.

The free parameters are the electron energy distribution index $p$, $\epsilon_e$, and $\epsilon_B$, with fixed parameters: kinetic energy $E_k = 1\times 10^{55}$~erg and circumburst density $n_0 = 1$~cm$^{-3}$, which have been measured relatively well or guessed reasonably. Other parameters such as its redshift $z=0.151$, and luminosity distance $D_L = 745$~Mpc. In addition, $\phi = 30$, where $\phi^{-1}$ corresponds to the acceleration efficiency of electrons, has been assumed. Extragalactic background light attenuation is implemented using the model of Saldana-Lopez et al.~(2021) \citep{saldana-lopez_observational_2021-1}. Flat priors are imposed with constraints: $1.8<p<2.5$ (excluding $1.95<p<2.05$ to avoid numerical instabilities), $10^{-3}<\epsilon_e<0.5$, $10^{-4}<\epsilon_e<0.5$ and $\epsilon_e>\epsilon_B$. The log-likelihood is computed from a $\chi^2$ statistic that accounts for asymmetric measurement errors, with the joint log-probability of two time interval. The result of fitting with LHAASO's and AGILE's data as shown in Fig.~\ref{fits2}.  Figure~\ref{MCMC2} shows the corner plot of the MCMC fitting. We get the best-fit parameters with $1\sigma$-confidence: $p_1 = 2.053_{-0.003}^{+0.034}$, $\epsilon_e = 0.0021_{-0.0006}^{+0.0001}$, and $\epsilon_B = 0.0006_{-0.0001}^{+0.0001}$.

\begin{figure}[htbp]
  \centering
  \includegraphics[scale=0.4]{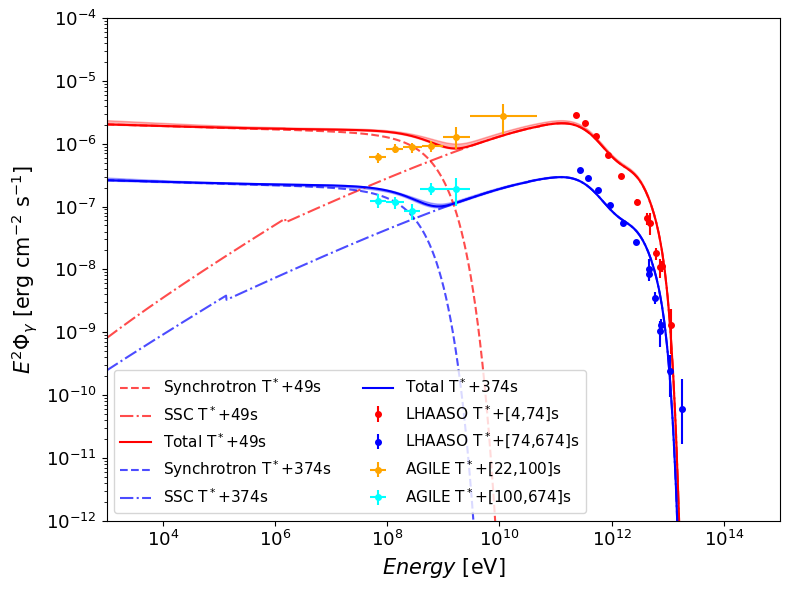}
  \caption[]
  {\label{fits2}{SED of GRB~221009A from the SSC model fit to the LHAASO (KM2A and WCDA) and AGILE data in two time intervals. The MCMC 1$\sigma$-confidence band is shown over the model for the best-fit parameter reported in the main text.}}
\end{figure}

\begin{figure}[htbp]
  \centering
  \includegraphics[scale=0.48]{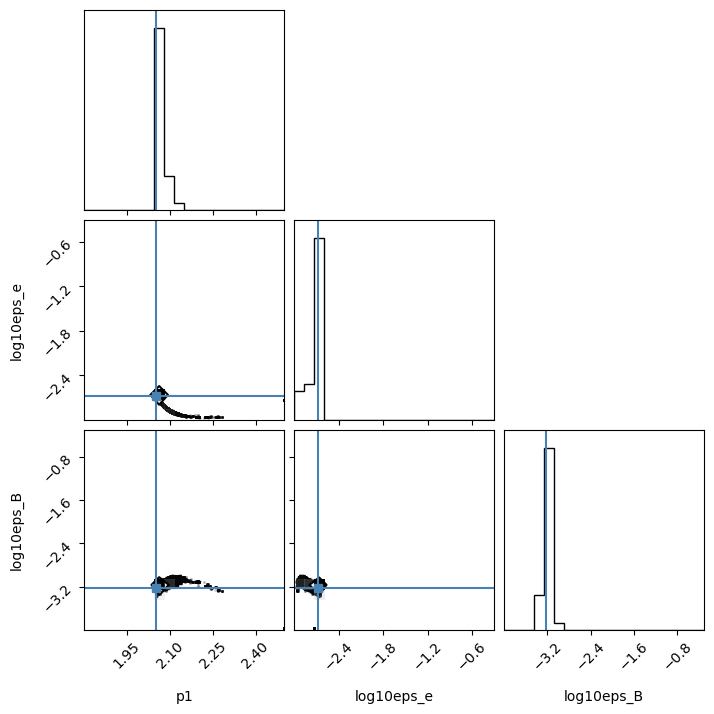}
  \caption[]
  {\label{MCMC2}{Corner plot showing the posterior distributions of the SSC model parameters obtained from an MCMC fit to the LHAASO and AGILE data of GRB~221009A in two time intervals as in Fig.~\ref{fits2}.}}
\end{figure}

\section{Details of detector configurations}\label{app1}
\begin{figure}[htbp]
  \centering
  \includegraphics[scale=0.55]{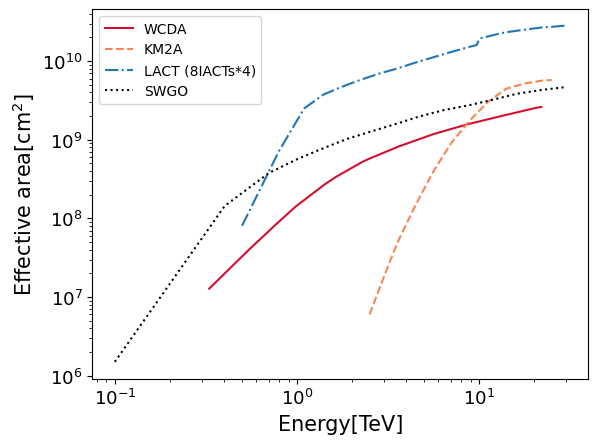}
  \caption[]
  {\label{A_psf}{Effective area after p/$\gamma$ discrimination for LHAASO, LACT (32 IACTs) and SWGO . The red solid curve shows the gamma-ray effective area of WCDA; the orange dotted curve corresponds to KM2A \citep{the_lhaaso_collaboration_very_2023}. The blue dash-dotted curve represents the collection area of LACT with 32 IACTs \citep{zhang_layout_2025}, and the black dotted line indicates the effective area of SWGO \citep{collaboration_science_2025-1}.}}
\end{figure}

The effective area and gamma-ray survival fraction for LHAASO are obtained from private communication with the LHAASO collaboration. For LACT, we scale the reported effective area of the existing 8‑IACT array by a factor of four to approximate the performance of the planned 32 IACT configuration \citep{zhang_layout_2025}. For SWGO, we adopt the effective area corresponding to its latest reference design \citep{collaboration_science_2025-1}.

The background for all instruments is dominated by cosmic rays. We generate the background event counts $N_{\rm bkg}$ using the cosmic‑ray spectra of five nuclei (H, He, CNO, Mg-Si, and Fe) from the Hillas model (H3a) \citep{gaisser_cosmic_2013}.

Event numbers are calculated via Eq.~\ref{count}. The effective area $A(E)$, survival fraction $\epsilon(E)$, and energy bins differ for each observatory. We compute the source and background events separately for WCDA, KM2A, LACT, and SWGO using the effective areas after p/$\gamma$ discrimination shown in Fig~\ref{A_psf}. The number of events within the source region is defined by the 68\% containment radius of the point‑spread function (PSF) \citep[]{aharonian_performance_2021, zhang_prospects_2024-2, collaboration_science_2025-1}. Further details are provided below.


\noindent\textbf{WCDA.} The effective area at a zenith angle of $30^\circ$ is adopted for both protons and gamma-ray. We apply the survival fractions $\epsilon_p(E)$ and $\epsilon_\gamma (E)$ to obtain $N_{\rm src}$ and $N_{\rm bkg}$ after discrimination. The energy bin edges are [0.33, 0.75, 1.13, 1.92, 3.27, 5.33, 11.42, 22]~TeV.

\noindent\textbf{KM2A.} The effective area at $30^\circ$ zenith is multiplied by fixed survival fractions $\epsilon_\gamma = 0.74$ and $\epsilon_p = 0.02$ to compute $N_{\rm src}$ and $N_{\rm bkg}$ via Eq.~(\ref{count})\citep{the_lhaaso_collaboration_very_2023}. The logarithmic bin edges are $\log(E_{\rm bin}/\rm TeV) = [0.4,0.6,0.8,1.0,1.2,1.4]$

\noindent\textbf{LACT.} We adopt the effective area and 68\% containment angular resolution for the 400~m setup within a group of IACTs \citep{zhang_layout_2025}. The survival fraction is taken {from}~\citep{zhang_prospects_2024-2}. Because the published collection area corresponds to an 8‑telescope array, we optimistically multiply it by four to estimate the effective area (before discrimination) of the final 32‑telescope configuration. The energy range is divided into nine bins spanning 0.5-30~TeV.
KM2A’s excellent particle discrimination capability \citep{zhang_prospects_2024-2} can significantly enhance LACT’s sensitivity in a joint observation. We assume all events detected by LACT can pass KM2A’s preselection. For the survival fraction of LACT, we adopt the value at 10~TeV for the 0.5–10~TeV range, and the KM2A discrimination values for energies above 10~TeV.

\noindent\textbf{SWGO.} We use the reported effective area for the full array at zenith angles between $0^\circ$ and $30^\circ$ \citep{collaboration_science_2025-1}. The angular resolution and survival fraction are taken from the best‑estimate values across all operational zones.

\end{document}